# Design of an RSFQ Control Circuit to Observe MQC on an rf-SQUID

Roberto C. Rey-de-Castro, Mark F. Bocko, Andrea M. Herr, Cesar A. Mancini, Marc J. Feldman

*Abstract*—We believe that the best chance to observe macroscopic quantum coherence (MQC) in an rf-SQUID qubit is to use on-chip RSFQ digital circuits for preparing, evolving and reading out the qubit's quantum state. This approach allows experiments to be conducted on a very short time scale (sub-nanosecond) without the use of large bandwidth control lines that would couple environmental degrees of freedom to the qubit, thus contributing to its decoherence. In this paper we present our design of an RSFQ digital control circuit for demonstrating MQC in an rf-SQUID. We assess some of the key practical issues in the circuit design including the achievement of the necessary flux bias stability. We present an "active" isolation structure to be used to increase coherence times. The structure decouples the SQUID from external degrees of freedom, and then couples it to the output measurement circuitry when required, all under the active control of RSFQ circuits.

## INTRODUCTION

THE goal of the experiment described in this paper is to demonstrate a quantum superposition state in a macroscopic superconducting circuit, the rf-SQUID. Superconductors offer a promising arena to observe quantum coherent behavior because at temperatures well below their transition temperature superconductors have very low entropy, which indicates that there are very few available internal modes that can contribute to decoherence. In the rf-SQUID, if an external magnetic flux bias equal to $\phi_0/2$, where $\phi_0 = 2.07 \times 10^{-15}$ V-sec, is applied and if $1 < 2\pi L I_c/\phi_0 < 5\pi/2$, where L is the inductance of the rf-SQUID and $I_c$ is the critical current of its Josephson junction, then the potential energy function is a symmetric double well [1]. The potential as a function of the total flux $\phi$ linking the SQUID is

$$V(\phi) = (\phi - \phi_x)^2/2L - (I_c \phi_0/2\pi) \cos(2\pi\phi/\phi_0), \quad (1)$$

where $\phi_x$ is the externally applied magnetic flux bias. The Hamiltonian of the system is given by [2]

$$H = -(\hbar^2/2C) \, d^2/d\phi^2 + V(\phi), \quad (2)$$

where C is the junction capacitance. The eigenvalues and eigenfunctions of the Hamiltonian must be found by numerical solution. With suitably chosen parameters (see below), the two lowest energy levels lie close to the bottom of the

Manuscript received September 18, 2000. This work was supported in part by ARO grant # DAAG55-98-1-0367.
The authors are with Electrical and Computer Engineering Department, University of Rochester, Rochester, NY 14627. E-mail: sde@egroups.com.
C. A. Mancini was with the Rochester ECE Department. He is now with Motorola Inc., Semiconductor Products Sector, Architecture and System Platforms, Austin, TX 78729. E-mail: Cesar.Mancini@Motorola.com.

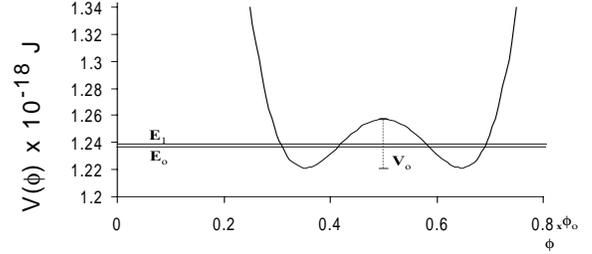

Fig.1. Potential energy and the two lowest energy levels in the rf-SQUID biased at $0.5\phi_0$.

potential and are well separated from the other, higher energy levels. Thus the rf SQUID may be approximated as a two level system (see Fig. 1).

The eigenfunctions corresponding to these energy levels, $\varphi_0$ and $\varphi_1$, are respectively even and odd functions of the flux centered on $\phi = 0.5\phi_0$. The linear combinations of these two energy states, $\Psi_\pm = 1/\sqrt{2} \, (\varphi_0 \pm \varphi_1)$ are functions centered on the left ($\Psi_+$) or the right ($\Psi_-$) wells and correspond to a current in the rf SQUID of approximately $0.8 I_c$ circulating counter-clockwise for $\Psi_+$ and clockwise for $\Psi_-$. If at $t = 0$ the system is localized in the left well, the probability of finding it in the right well after a time t is

$$P_{+-} = \tfrac{1}{2}\,[1 - \cos((E_1 - E_2)t/\hbar)] . \quad (3)$$

If one experimentally prepares the system to be in the left well at $t = 0$ and then measures the probability as a function of time of finding the system in the right well the observation of the oscillation predicted in (3) would provide the sought-for demonstration of macroscopic quantum coherence.

In order to ensure that we are dealing with a two level system, our two levels should be isolated from the other levels, so they must lie well below the potential barrier and the thermal occupation of the higher states should be negligible. This requires $k_B T \ll \hbar\omega_p < V_0$, where $\omega_p$ is the junction plasma frequency and $V_0$ is the height of the barrier (see Fig. 1). Furthermore, to prevent thermalization of the quantum state composed from the two low lying energy levels, we require: $k_B T/Q < \Delta E = E_1 - E_0$ where Q is an effective "quality factor" of the rf-SQUID. Q is the number of cycles of oscillation at the SQUID plasma frequency, $\omega_p = 1/(LC)^{1/2}$, over which the SQUID loses "memory" of its quantum state, i.e., suffers decoherence. $Q = (1/2\ln 2)\,\omega_p t_d$, where $t_d$ is the assumed decoherence time.

This allows us to estimate the necessary temperature to be able to observe MQC in our proposed experiment. For a coherence time of $t_d = 10$ nsec, $C = 50$ fF, $L = 97$ pH and $I_c = 3.8$ μA, we have $\omega_p \approx 450$ GHz and $Q \approx 3 \times 10^5$. For this Q value $k_B T/Q = 1.4 \times 10^{-29}$ J at $T = 0.3$ K, and this is more than a factor of 10 smaller than the separation of the lowest two energy levels even for a MQC frequency of 1 MHz.

We believe that the scheme we propose has a good chance of success despite the failure of related experiments attempted by other groups [3]-[6]. Previous experiments that attempted to observe MQC oscillations in an rf-SQUID chose typical parameters as follows $I_c$ = 50 nA, C = 0.2 pF, and L = 10 nH [5], yielding oscillation frequencies on the order of 10 MHz. A successful demonstration would require the maintenance of quantum coherence for several microseconds, requiring excellent isolation of the SQUID from external (and internal) degrees of freedom which is especially difficult given that large bandwidth signal carrying lines were used to couple the SQUID to external room temperature electronics. For our parameters [7] the oscillation frequency is 1.12 GHz, and we intend to measure the oscillation using very fast on-chip RSFQ digital circuits that allow us to both isolate the rf SQUID qubit better and perform the measurement faster than previous attempts.

OVERVIEW

The goal of the proposed experiment is to measure the probability of finding the flux expectation value of the rf-SQUID centered in the right hand well of the potential (1) at time t, after starting with the system localized in the left well at time t = 0. In order to accomplish this, we must prepare the rf-SQUID flux in the left-hand well and make a measurement at a chosen time in the future to determine in which well the SQUID is found. The process of state preparation, evolution for time t, and measurement will be repeated many times in order to determine the probability of finding the SQUID in the right well. The measurement will be repeated for various time intervals t, in order to find the probability of occupying one well or the other as a function of time.

Accordingly, the control circuit must bias the SQUID at precisely $\phi_0/2$, prepare the SQUID in its initial state, and read out the SQUID state after a chosen time interval. We achieve these functions using fast RSFQ digital circuits that have only low speed connections to the outside world. From the outside we first send a number of pulses that set the time interval t, then a SET pulse is sent to initially localize the flux on the left well and finally a START pulse is sent to begin the experiment. At the end of the experiment, an output pulse is obtained only if the system has its flux located in the right-hand well at time t.

First, we will explain in detail how the control circuit achieves its goals. Then we will explain the schemes that we use to maintain qubit coherence in the face of the interactions with its RSFQ control circuit.

THE CONTROL CIRCUIT

A. Flux Bias

The required flux bias is provided to the qubit by an inductively coupled high critical current rf-SQUID (Fig. 2). When the current bias to its junction is increased, fluxoids can enter the rf-SQUID one by one. When the current bias is turned off, the fluxoids stay in the loop and a current close to that required to produce the required flux bias circulates permanently through the SQUID inductor. The current is fine-tuned to its precise required value by changing $I_{extra}$.

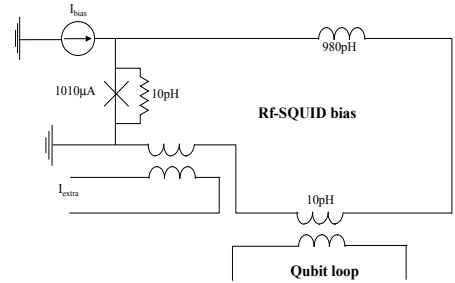

Fig. 2. The rf-SQUID flux bias scheme.

B. Timing

The main part of the control circuit is dedicated to providing the correct timing between the beginning and the end of the experiment. One advantage of using RSFQ digital circuits is that they work at very high frequencies; for example, using a 10 GHz clock will allow us to make measurements with a time resolution of 0.1 nsec.

First, we must have the ability to set the rf-SQUID ("qubit loop") in state "0", in which the flux is localized in the left hand potential well. Then after a programmable time interval we wish to measure the flux in the SQUID to determine if it is in the "1" or "0" state.

The diagram of the experiment is shown in Fig. 3. We set the rf-SQUID in state "0" by decreasing the current on the "rf-SQUID bias" from the value that gives exactly $\phi_0/2$ external flux bias in the qubit loop. That is accomplished by storing a fluxoid in a "Tipping" DRO (destructive read-out) that is inductively coupled to the rf-SQUID bias loop. To accomplish this a SET pulse is sent to the DRO "in" port. The SET pulse is also sent to the "in" port of the END DRO so that it can be ready to receive the output from the TFF (toggle flip-flop) chain to produce the OUT pulse. The SET pulse is also sent to the "in" port of the "biased dc-SQUIDs" 2 and 3, which sets them in their high inductance state.

A 10 GHz Josephson ring oscillator is connected to a chain of 7 TFF's in series to set the time interval. The TFF chain output period is longer than the input by a factor of $2^7$=128. So, by setting the initial bits in the TFF one can achieve any desired delay from 0.1 nsec to 12.8 nsec. The ring oscillator contains an independently biased switch, which works because it only allows SFQ pulses to circulate when its bias exceeds a threshold value; otherwise the pulse escapes through an escape junction. With the switch opened we send a train of n pulses to the TFF chain. Then, we close the switch and send one "START" pulse to the ring oscillator. The TFF chain will produce an output after a time

$$t = (128-n) T$$

where T is the period of the ring oscillator. After this pre-programmed time the TFF output pulse is produced and reaches the clock input of the "END" DRO and releases the previously stored SET pulse to produce the OUT pulse. After the first TFF output pulse many others are produced, but they arrive at the clock input of the empty "END" DRO and will not produce any further output, but will simply escape through the escape junction at the DRO clock input. After op-eration there is no way to know the state of the TFF chain. But, the TFF chain can be emptied at the end of the experiment by sending a low speed train of pulses until one output is detected in the TFF out SFQ/dc converter. The TFF chain can be readied for the next experiment in this manner.

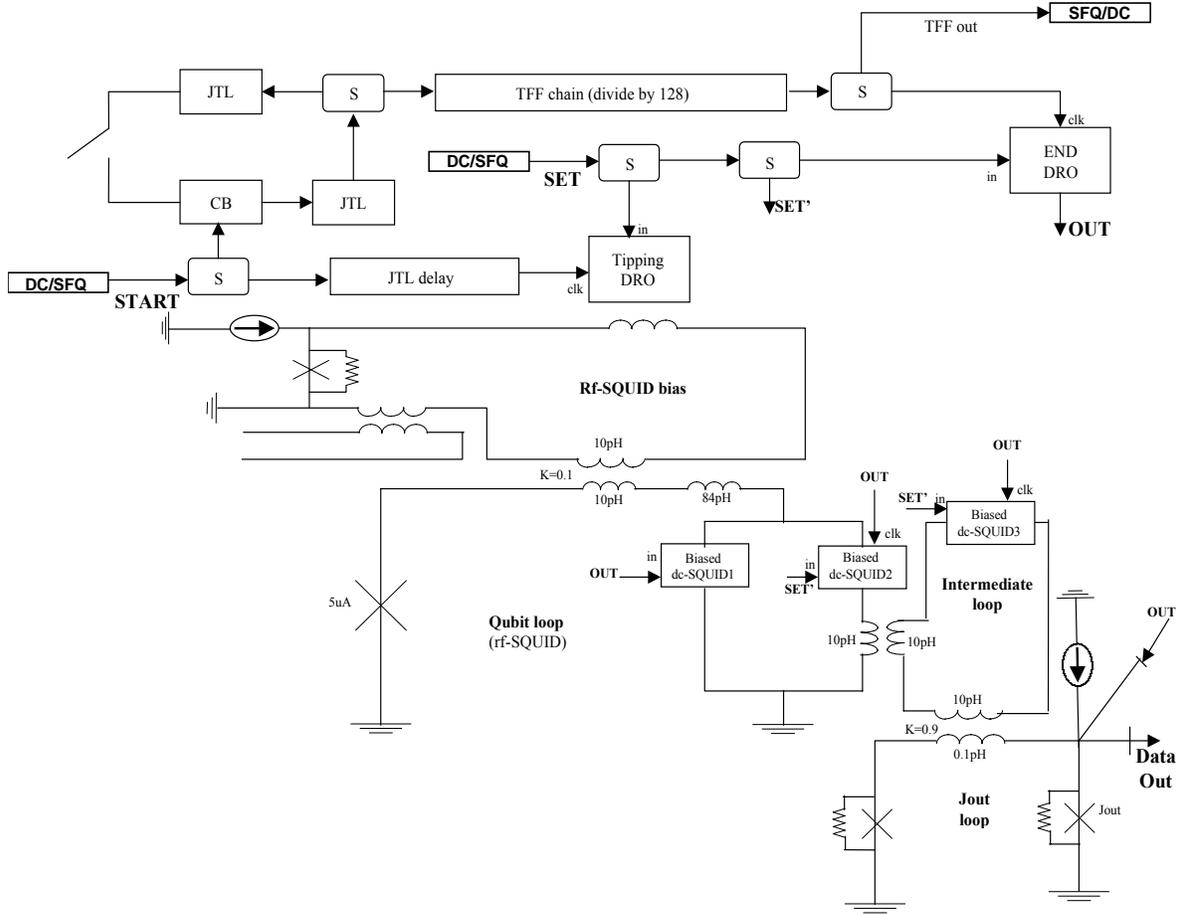

Fig. 3. Diagram of the control circuit. "S" represents a splitter which gives two separate output SFQ lines for a single input line. "CB" represents a confluence buffer which allows separate input SFQ lines to feed the same output line. The other circuit blocks are explained in the text.

The experiment begins when the START pulse arrives at the clock input of the "tipping" DRO, releasing the trapped fluxoid (that was there since the arrival of the SET pulse) and restoring the external flux bias on the qubit loop to $\phi_0/2$, thus commencing the free evolution of the qubit. The START pulse arrives at the "tipping" DRO only after passing a JTL (Josephson transmission line) delay line shown in the diagram. The JTL delay is chosen to compensate for the delays in the OUT pulse path, to make the time between the beginning and the actual measurement of the state of the qubit loop equal to the pre-programmed time. Simultaneously, the START pulse is sent to the ring oscillator with its switch closed, so that the ring oscillator begins to feed the TFF chain with a high frequency train of pulses eventually producing an output from the TFF chain after a set time interval. This output will serve to read-out the state of the qubit loop.

### C. Measurement

We have described the way to begin and end the experiment; in what follows the measurement process will be described. The output circuitry shown in Fig. 3 is considerably more complicated than that suggested in Ref. [7], in an attempt to more completely isolate the qubit from the RSFQ measurement circuitry. Each biased dc-SQUID element consists of a dc-SQUID inductively coupled to a DRO (see Fig. 4). When the DRO stores a fluxoid, it is in the "high" i.e. high inductance state. During the qubit free evolution, the biased dc-SQUIDs 2 and 3 are in their high states while the biased dc-SQUID 1 is in its "low" i.e. low inductance state (no current in its biasing DRO). In this way only a small part of the current in the qubit loop goes through the biased dc-SQUID 2 path, which contains the only inductance that is coupled to the "intermediate loop". This makes the induced flux in the intermediate loop very small and an accordingly small current flows in this loop. In turn this makes the current in the loop containing $J_{out}$ very small. When the OUT pulse arrives it enters the DRO of the biased dc-SQUID 1 through its "in" port, setting it in its high state, and releases the pulses stored in the biased dc-SQUIDs 2 and 3, setting them to their low state. In this way the situation is reversed and the current induced in the intermediate loop is much greater than before, which produces a greater current in the $J_{out}$ loop. Simultaneously, the OUT pulse has reached $J_{out}$ in the loop contiguous to the $J_{out}$ loop, producing a current on $J_{out}$ that is barely under its critical current. Note that the biased dc-SQUID 2 has to be somewhat different from the other two in order not to influence in the dynamics of the qubit loop. The critical currents of both junctions in its dc-SQUID are larger than the other biased dc-SQUIDs. One limitation of this scheme [8] may be the difficulty to realize the small values required for the dc-SQUID geometrical inductances (Fig. 4).

As mentioned in the introduction, the two localized linear combinations $\Psi_+$ and $\Psi_-$ in the qubit correspond to current in the rf-SQUID flowing in opposite directions. When the qubit is in the $\Psi_+$ state "0" the current on $J_{out}$ will be reduced and when it is in state $\Psi_-$ "1" the current on $J_{out}$ will be increased. The parameters in the $J_{out}$ loop have been chosen so that the slight increase in current when the qubit is in state "1", will make $J_{out}$ "flip" producing a fluxoid that is detected at the "Data Out" SFQ/dc converter. This happens only if the qubit is in state "1".

## QUBIT ISOLATION

RSFQ circuits provide an excellent barrier between the qubit and the environment. Each junction to ground provides some isolation. A small signal coming from either side would be shorted by the junction, because for small signals the junctions are approximately equivalent to inductors that are, in most cases, small. However a large signal such as an SFQ pulse may make its way through a series of junctions as a solitary wave; this is because of the nonlinearity of the junction's response to large versus small excitations. Although external modes are completely shielded by the SFQ circuits, the shunt resistors required for the SFQ circuit junctions are themselves a serious concern.

The control circuit can interact with the qubit via the "rf-SQUID bias" loop and the "Intermediate" loop.

The parameters of the rf-SQUID bias (Fig. 2) are chosen so that the mutual inductance between it and the qubit loop is much smaller than the total inductance of either loop. The induced current in the qubit loop from the rf-SQUID bias loop is typically $10^{-5}$ of the current of the qubit loop. It is possible to reduce the mutual inductance to the rf SQUID bias loop because we can always increase the current in it and, in fact, it is designed to support a huge number of fluxoids.

This weak-coupling approach is not possible in the read-out part of the circuit, because it should be able to detect the small currents produced in the qubit loop. Instead, as described in the previous section, we designed an isolation structure in which the output coupling is actively controlled by SFQ pulses. The measurement circuitry is very well isolated from the qubit during the quantum coherent evolution of the qubit wavefunction. Then when it is time for a (classical) measurement to be made, the measurement circuitry is abruptly switched into closer electrical contact with the qubit.

In a future publication we will present various "passive" structures ("filters") for isolating quantum coherent circuitry from RSFQ circuits. These do not need to be switched in and out of play using SFQ pulses.

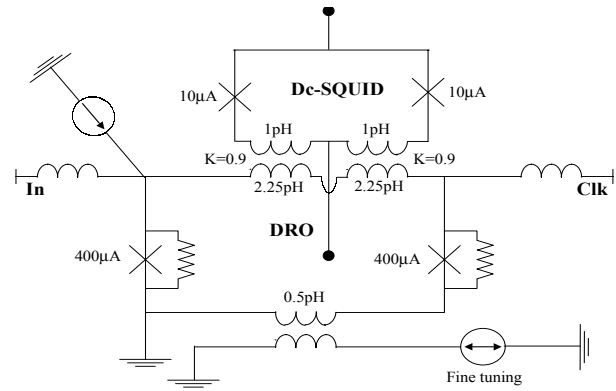

Fig. 4. The "biased dc-SQUID" element, which provides active control of the output coupling.

## CONCLUSION

We have presented our design for an RSFQ digital control circuit for demonstrating MQC in an rf-SQUID. We have discussed many of the key practical issues for this experiment. We present an "active" isolation structure to be used to increase coherence times. The structure decouples the SQUID from external degrees of freedom, and then couples it to the output measurement circuitry when required, all under the active control of RSFQ circuits.

## ACKNOWLEDGMENT

We are grateful to Pavel Rott, Xingxiang Zhou, and John L. Habif for helpful discussions.